%% file: stf_icml_forget.tex
\documentclass{article}

\usepackage{microtype}
\usepackage{graphicx}
\usepackage{subfigure}
\usepackage{booktabs} 

\usepackage{hyperref}

\usepackage{threeparttable}
\usepackage{amsmath}
\usepackage{forest}
\usepackage{amssymb}
\usepackage{mathtools}
\usepackage{amsthm}
\usepackage{xypic}
\usepackage{algorithm}
\usepackage{algorithmic}
\usepackage[capitalize,noabbrev]{cleveref}

\theoremstyle{plain}
\newtheorem{theorem}{Theorem}[section]

\theoremstyle{definition}

\theoremstyle{remark}

\renewcommand{\algorithmicrequire}{\textbf{Input:}}
\renewcommand{\algorithmicensure}{\textbf{Output:}}
\usepackage[textsize=tiny]{todonotes}

\usepackage{subfiles}
\usepackage{pgfplots}

\begin{document}

\title{Morse-STF: Improved Protocols for Privacy-Preserving Machine Learning}

\author{Qizhi~Zhang, Sijun~Tan, Lichun~Li, \\
Yun~Zhao, Dong~Yin, Shan~Yin \\
\{qizhi.zqz, sijun.tsj, lichun.llc, caesar.zy, \\
yindong.yd, yinshan.ys\}@antgroup.com
     \thanks{The first two authors contributed equally, and they share the co-first authorship.}
}



\maketitle

\begin{abstract}

Secure multi-party computation enables multiple mutually distrusting parties to perform computations on data without revealing the data itself, and has become one of the core technologies behind privacy-preserving machine learning. 
In this work, we present several improved privacy-preserving protocols for both linear and non-linear layers in machine learning. For linear layers, we present an extended beaver triple protocol for bilinear maps that significantly reduces communication of convolution layer. For non-linear layers, we introduce novel protocols for computing the sigmoid and softmax function. Both functions are essential building blocks for machine learning training of classification tasks. Our protocols are both more scalable and robust than prior constructions, and improves runtime performance by 3-17x. Finally, we introduce Morse-STF, an end-to-end privacy-preserving system for machine learning training that leverages all these improved protocols. Our system achieves a 1.8x speedup on logistic regression and 3.9-4.9x speedup on convolutional neural networks compared to prior state-of-the-art systems. 

\end{abstract}

\section{Introduction}

Machine learning (ML) applications are "data-hungry", as more data usually leads to better concrete performance of machine learning models. The growing amount of data required to train a machine learning model also raises significant privacy concerns. For example, multiple organizations may want to jointly train a machine learning models by aggregating their data, but want to keep their data private to the other parties for privacy reasons. Secure multi-party computation (MPC) \cite{MPC1, MPC2} is a cryptographic primitive that can achieve the above objective. In secure MPC, parties can jointly obtain the output of the models without revealing their input data to the other parties. Recently, there has been an active line of research in combining machine learning with MPC to achieve privacy-preserving machine learning (PPML). (e.g. \cite{SecureML, ABY3, SecureNN, ABY2, cryptflow, CrypTFlow2, TFE, Rosetta, crypten2020, Falcon, Delphi, TKTW21})


Most of these work focus on machine learning inference, and only a few considers machine learning training~\cite{SecureNN,Falcon,crypten2020,TKTW21}. Compared to machine learning inference, machine learning training is much more computationally demanding. First, inputs of machine learning training usually comes in large batches, whereas machine learning inference has a batch size of one. Second, machine learning training requires computing an extra back propagation step over the network to update the network parameters. The back propagation of the loss function typically involve non-linear operations that are expensive to compute privately. The large input batch size and the back propagation step both significantly increase computation and communication between parties to execute MPC protocols, and make privacy-preserving ML training a challenging task.


In this work, we present several novel protocols that significantly reduce the computation and communication burden of privacy-preserving ML training. Our protocols are based on new approaches that are not present in the MPC literature, and therefore shed light on more efficient protocol design for other ML operations. 

\textbf{Our  contribution.} The main contribution of our work can be summarized as follows:
\begin{itemize}
    \item We propose an extended beaver triple protocols for bilinear maps, and prove that both the forward and backward propagation of convolution forms a bilinear map. By using the extended beaver's protocol on convolution, we reduce the runtime of convolution's back propagation on filters by 5.4-9.1x.  
    \item We propose a novel protocol to compute the sigmoid function based on Fourier series. Compared to prior approach that approximate sigmoid using the Chebyshev polynomial, our protocols reduces online round complexity by 80\%, and online communication by 97\%.
    \item We propose a novel protocol for computing softmax based on initial value problems of ordinary differential equations. Our protocol only requires a constant 32 rounds, and produces robust output even when the class size is 10000. Compared to prior constructions, our protocol achieves a 3-17x runtime speedup on different class sizes. 
    \item We build Morse-STF, a system for privacy-preserving ML training built upon Tensorflow~\cite{tensorflow2015-whitepaper} that leverages these new protocols. Overall, our system achieves a 1.8x runtime speedup on logistic regression training, and a 3.9-4.9x speedup on convolutional neural network training compared to prior state-of-the-art systems.
\end{itemize}

\subfile{mpc}

\subfile{bilinear_map}

\subfile{non_linear}

\subfile{experiment}


\subfile{conclusion}

\newpage



\bibliographystyle{alpha}
\bibliography{paper.bib}

\newpage
\appendix

\subfile{appendix}


\end{document}

%% file: mpc.tex
\section{Secure Multi-Party Computation}
In this section, we will first discuss the threat model we use in our system, and then provide some preliminaries on secure multi-party computation.  

\subsection{Threat Model}
Similar to many prior works~\cite{ABY3,cryptflow, crypten2020, Falcon}, we design our system in the 3-party setting with a focus on semi-honest adversaries. In this threat model, we assume that each party will honestly follow the protocols, but may try to learn information about other parties input. We refer readers to these prior works for a formal definition of 3PC semi-honest security.

Our work assumes there is a trusted third-party (TTP) to assist offline pre-processing phase. The TTP assist the other two parties by sending them resources, but do not engage in the actual computation of the cryptographic primitives. This setting is captured by the notion of commodity MPC~\cite{taas}, and is different from a recent line of work~\cite{ABY3, Falcon, TKTW21} that leverages 2-out-of-3 replicated secret sharing to reduce online communication. In these works, the third party actively involves in the joint computation, and can both send and receive shares from other parties. The ability for the third party receive shares from other two parties give the protocol designer more flexibility to design more efficient protocols.

Although both approaches belongs to the same 3-party threat model from a security standpoint, in practice these two approaches makes a non-trivial difference. In a practical scenario, the third party is often also the MPC-service provider. The underlying MPC algorithms are often times black-box to the other two parties, so they do not trust an MPC provider that receive any shares from them. (the service provider can easily tamper with the algorithm to receive their data in plaintext). In contrast, a third party that only assist computation by sending them resources requires less trust, and are much more acceptable (the third-party alone can not recover the plaintext data since it receive nothing). Moreover, in many cases, we can offload the resource generation of the third party to the offline phase and turn the online phase into a 2-party setting. In our system, all resources generated by the third party can be preprocessed during the offline phase.

Since our work considers the commodity MPC setting, we only compare our system with prior 3PC work that has the same setup~\cite{SecureNN, TFE, Rosetta, crypten2020}. The new approaches we proposed for non-linear protocols are nevertheless independent of the underlying setting and can be extended to improve performance for more general MPC settings.

\subsection{Secret Sharing \& Beaver Triples}
In MPC, inputs to the function are secret-shared among the parties so that no single party gets to see the true inputs. There are two types of secret-sharing that is mainly used in privacy-preserving machine learning: arithmetic secret-sharing and binary secret-sharing.

In arithmetic secret-sharing, a scalar value $x \in \mathbb{Z}_N$ is secret-shared between 2 parties so that $x = [x]_0 + [x]_1 \mod N$, where party $i$ holds $[x]_i$. Addition of two scalar values $x, y$ can be evaluated locally by locally summing up the shares $[x]_i + [y]_i$. For multiplication, the TTP generates beaver triples $a * b = c \mod N$ where $a, b, c$ are randomly sampled from $\mathbb{Z}_n$ and distribute $[a]_i, [b]_i, [c]_i$ to each party. The two parties then engage in the beaver triple multiplication protocol~\cite{beaver} to obtain $[z]_i = [x * y]_i$.

In binary secret-sharing, a scalar value $x$ is decomposed into bits where each bit $x_i \in \{0, 1\}$ is secret-shared as $x_i = [x_i]_0 \bigoplus [x_i]_1 \mod 2$. The XOR gate in binary sharing is analogous to addition in arithmetic secret sharing, and requires no communication. The AND gates is analogous to multiplication in arithmetic sharing, which requires the TTP to distribute corresponding triples. 

Arithmetic secret-sharing is suitable for the majority of linear operations in machine learning (e.g. matrix multiplication, convolution) whereas non-linear operations such as ReLU are more efficiently computed using binary secret-sharing. Protocols to convert between arithmetic and binary shares has been studied extensively~\cite{ABY, ABY3, ABY2}. ReLU is the only component in our system that requires share conversion. Our ReLU protocol follows directly from prior approach and has similar communication complexity. We therefore do not discuss the ReLU protocol in our paper, and we refer readers to~\cite{ABY3} for a detailed description of the approach. The main contribution of our work comes from novel protocols to evaluate sigmoid and softmax that removes the expensive share conversion step. We describe these protocols in more detail in Section~\ref{sec:non-linear}.

%% file: bilinear_map.tex
\section{Bilinear Map for Linear Layers}
\label{sec:bilinear}
The major computation in machine learning is linear layers such as matrix multiplication and convolution. Linear layer consists of a series of addition and multiplication, and since addition can be computed locally, all communication comes from multiplication, which requires the TTP to generate and distribute the beaver triple. 

The communication traffic does not necessarily scale linearly with the number of multiplications in a linear operation. For certain types of linear operations, we can leverage its mathematical structure to reduce the number of beaver triples sent. A simple example is matrix multiplication. To multiply two matrices $A \in M_{m,n}(\mathbb{Z}/N\mathbb{Z})$ and $B \in M_{n,p}(\mathbb{Z}/N\mathbb{Z})$ to obtain $C \in M_{m,p}(\mathbb{Z}/N\mathbb{Z})$, parties do not communicate $O(m * n * p)$ elements. Instead, the TTP generate random triple of matrices $R_1R_2=R_3$ that has the same shape as $A, B, C$ respectively, which only has $O(m*n + n*p + m*p)$ elements to communicate.

This approach to generate beaver triple is in fact commonly used in prior literature ~\cite{Falcon, crypten2020,cryptflow} to perform matrix multiplication. However, it is not obvious to see that we can generalize this approach to other linear operations such as convolution. Many prior works~\cite{SecureNN, Falcon} in privacy-preserving ML performs convolution by converting the convolutional layer into its equivalent matrix form and performing matrix multiplication instead. This approach incurs non-trivial communication overhead since the resulting matrix is much larger than the original convolution kernel. In fact, we can distribute beaver triples for convolution more efficiently as well. 

We therefore introduce and formalize the mathematical notion of bilinear map. We present a beaver triple protocol for bilinear map that is a natural generalization to the original beaver triple protocol for scalar multiplication. Moreover, we prove that both the forward and backward propagation of convolution satisfies the definition of bilinear map, and we can therefore use the generalized beaver triple protocol $\Pi_{\mathsf{BM}}^{f, ss}$ to distribute beaver triple more efficiently. 

\paragraph{Definition of Bilinear Map.}
Let $A, B, C$ be there Abelian group, a map $f : A \times B \longrightarrow C$ is called bilinear if
$$
\begin{array}{l}
 f(a_1 + ka_2, b) = kf(a_1, b) + f(a_2, b) \\
 f(a, b_1+kb_2) = f(a, b_1) + kf(a, b_2) 
\end{array}
$$ for any $a_1, a_2 \in A, b_1, b_2 \in B$ and $k \in \mathbb{Z}$.

It is easy to see that both the forward propagation of matrix multiplication and convolution satisfies the definition of bilinear maps. Now we will show in Theorem \ref{theorem:back_prop} that, if the forward propagation of a layer forms a bilinear map, its backward propagation is also a bilinear map. 

\begin{theorem}
\label{theorem:back_prop}
Let $A$ be a commutative ring with 1.
Let $X, Y, Z$ be affine spaces over $A$, $(x_i)_i$, $(y_i)_j$, $(z_k)_k$ are  coordinates of $X$, $Y$, $Z$ respectively.
Let $f : X \times Y \longrightarrow Z$ be a bilinear map,  $loss$ be a smooth function over $Z$. Then we have

a. $\frac{\partial loss}{\partial x_i}(x, y)$ is independent  of 
$x$, and is bilinear of $(\frac{\partial loss}{\partial z_k})_k$ and $y$;

b. $\frac{\partial loss}{\partial y_j}(x, y)$ is independent  of 
$y$, and is bilinear of $(\frac{\partial loss}{\partial z_k})_k$ and $x$;
\end{theorem}

We leave the proof of Theorem \ref{theorem:back_prop} to Appendix~\ref{append:conv}. Since both forward and backward convolution forms a bilinear map, we can use the following extended beaver triple multiplication protocol for bilinear map $\Pi_{\mathsf{BM}}^{f}$ to privately evaluate convolution with much fewer communication. Our protocol relies on pseudorandom function (PRF)~\cite{prf} to generate shared source of randomness between parties.


\begin{algorithm}[htbp]  \footnotesize
  \caption{Protocol $\Pi_{\mathsf{BM}}^{f}$  }
  \label{alg:BM_PRF}
  \begin{algorithmic}[1]
    \REQUIRE
      $P_0$, $P_1$ hold the shares of  $a=a_0+a_1 \in A$,  $b=b_0+b_1 \in B$. $P_0$ and $P_2$ holds PRF$_0$, $P_1$ and $P_2$ holds PRF$_1$.
    \ENSURE
      $P_0$, $P_1$ get the shares of $f(a, b)$ over $C$.
  \STATE $P_2$ generates  $\tilde{a}_0 \in A$, $\tilde{b}_0 \in B$, $\tilde{c}_0 \in C$ using PRF$_0$;
  generates $\tilde{a}_1 \in A$, $\tilde{b}_1 \in B$ using PRF$_1$
   and compute 
 $\tilde{a}:=\tilde{a}_0+\tilde{a}_1$, $\tilde{b}:=\tilde{b}_0+\tilde{b}_1$ and $\tilde{c}_1:=f(\tilde{a}, \tilde{b})-\tilde{c}_0$;
 \STATE  $P_0$ generates  $\tilde{a}_0 \in A$, $\tilde{b}_0 \in B$, $\tilde{c}_0 \in C$ using PRF$_0$; $P_1$ generates $\tilde{a}_1 \in A$, $\tilde{b}_1 \in B$ with PRF$_1$; $P_2$ sends $\tilde{c}_1$ to $P_1$;
 \STATE $P_0$ computes $\delta a _0:=a_0-\tilde{a}_0, \delta b_0:=b_0 - \tilde{b}_0$;
 \STATE $P_1$ computes $\delta a _1:=a_1-\tilde{a}_1, \delta b_1:=b_1 - \tilde{b}_1$;
 \STATE $P_0, P_1$ exchange $\delta a_0, \delta a_1, \delta b_0, \delta b_1$, and compute $\delta a:=\delta a_0+\delta a_1$ and $\delta b:= \delta b_0+\delta b_1$;
    \STATE $P_0$ computes $c_0:=f(\delta a, b_0) + f(\tilde{a}_0, \delta b) + \tilde{c}_0$, $P_1$ computes $c_1:=f(\delta a, b_1) + f(\tilde{a}_1, \delta b) +  \tilde{c}_1$;
    \STATE RETURN $(c_0, c_1)$.
  \end{algorithmic}
\end{algorithm}

Compared to the matrix multiplication approach, our protocol can save communication by up to $97.3\%$. It is worth noting that CrypTen also leverages the bilinear property of convolution, and uses $\Pi_{\mathsf{BM}}^{f}$ to compute the forward propagation of convolution. However, they do not extend this protocol to compute the backward propagation of convolution. Our protocol achieves a $82.5\%$ communication savings on convolution's backward propagation on filter. We leave the detailed analysis of communication complexity between the protocols in Appendix~\ref{append:conv}. Our microbenchmarks in Section \ref{sec:microbenchmarks} shows that our protocol $\Pi_{\mathsf{BM}}^{f}$ significantly improves concrete runtime performance for backward convolution.

%% file: non_linear.tex
\section{Improved Non-Linear Protocols} \label{sec:non-linear}
In this section, we present our novel protocols for computing sigmoid and softmax, both of which are important activation functions for machine learning training.

\subsection{Sigmoid}
The sigmoid function is commonly used in machine learning algorithms for binary classification tasks, and is a fundamental building block in logistic regression.
\[
\mbox{sigmoid}(x) = \frac{1}{1+e^x}
\]
There are two common approach to evaluate the sigmoid function privately. The first line of approach is to approximate sigmoid using Chebyshev polynomial of degree 9~\cite{TFE}:
\[
\mbox{sigmoid}(x) \approx a_0 + a_1 x + a_3 x^3 + a_5 x^5 +a_7 x^7 + a_9 x^9
\]
where
\[
\begin{array}{l}
    a_0 = 0.5, \\
    a_1 = 0.2159198015, \\
    a_3 = -0.0082176259, \\
    a_5 = 0.0001825597, \\
    a_7 = -0.0000018848, \\
    a_9 = 0.0000000072
\end{array}
\].

\begin{figure}[h]
\begin{minipage}[t]{0.5\linewidth}
 \includegraphics[scale=0.27]{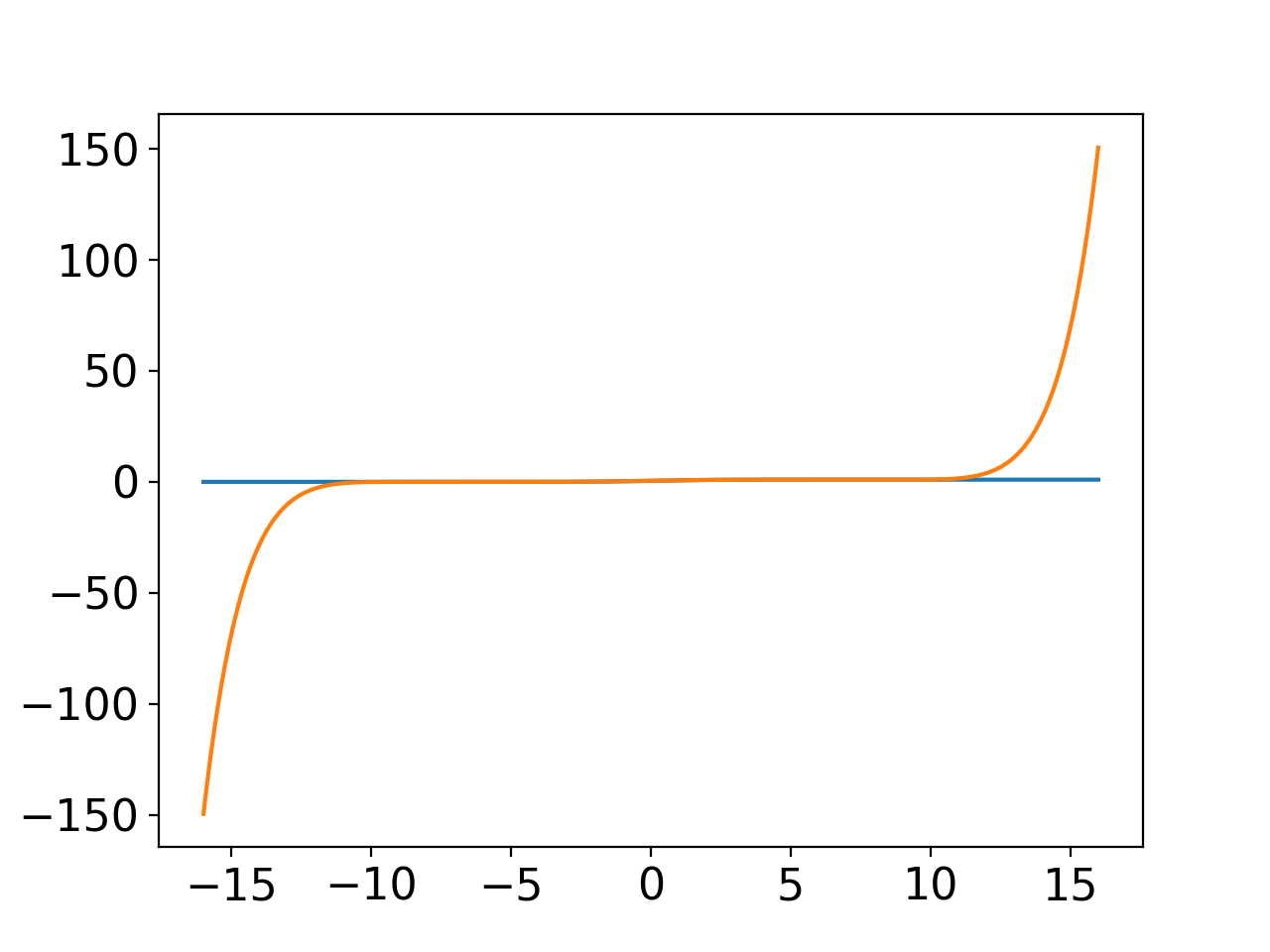}
\label{fig:sigmoid_poly}
\end{minipage}
  \begin{minipage}[t]{0.45\linewidth}
 \includegraphics[scale=0.27]{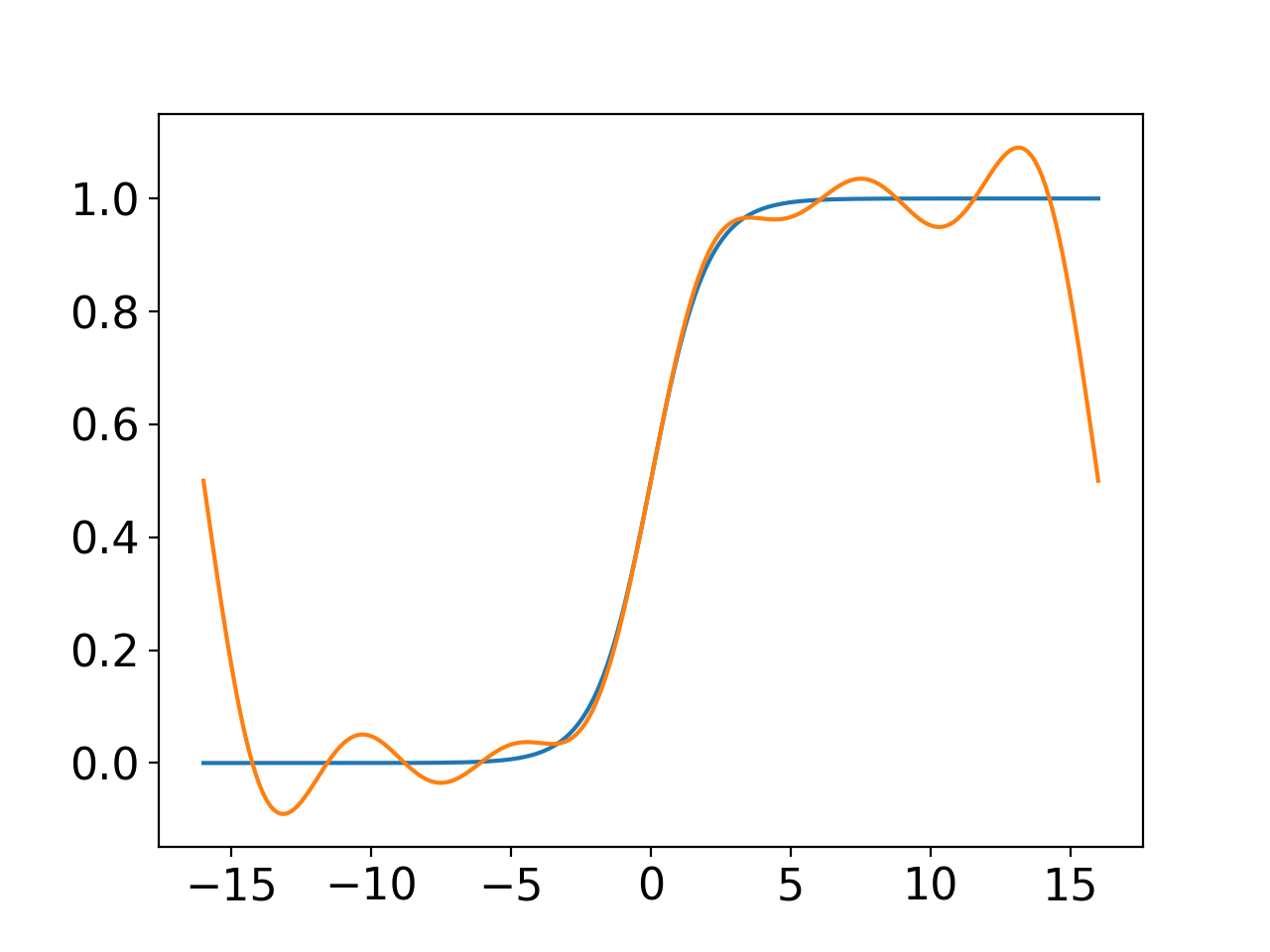}
\label{fig:sigmoid_fourier}
\end{minipage}
\caption{\textbf{Left}: sigmoid approximation with Chebyshev polynomial; \textbf{Right}: sigmoid approximation with Fourier series.}
\end{figure}

This approximation requires 5 rounds of communication. However, this approach only produces accurate approximation when the input $x$ is in the domain $(-8, 8)$. We plot the approximated sigmoid function using Chebyshev polynomial in the left sub-figure of Figure \ref{fig:sigmoid_poly}. It is easy to see that, when the input goes beyond $(-8, 8)$, the gradient starts to explode and produces results that are significantly off the chart.

Another line of approach \cite{ABY, ABY3, ABY2} approximate sigmoid using the following piecewise linear function:





\[
\mbox{sigmoid}(x) \approx \left\{
\begin{array}{ll}
0, & x<-1/2 \\
x+1/2, & -1/2 \leq x<1/2 \\
1 & x\geq 1/2 \\
\end{array}
\right.
\]
Although this approach does not experience exploding gradients, it requires private comparison which incurs a huge communication overhead, and the approximation is not as accurate.

In our work, we present a novel protocol to approximate sigmoid using Fourier series~\cite{fourier}:
\[
\begin{array}{rl}
\mbox{sigmoid}(x) \approx  & a_0 + a_1 \sin \frac{2\pi x}{32} + a_2 \sin \frac{4 \pi x}{32} +\\
  & a_3 \sin \frac{6 \pi x}{32} + a_4 \sin \frac{8 \pi x}{32} + a_5 \sin \frac{10 \pi x}{32}
 \end{array}
\] 
where
\[
\begin{array}{l}
  a_0=0.5 \\
  a_1= 0.61727893 \\
 a_2 = -0.03416704 \\
 a_3 = 0.16933091 \\
 a_4 = -0.04596946 \\
  a_5 = 0.08159136
  \end{array}
\]

From the right sub-figure of Figure \ref{fig:sigmoid_fourier}, the Fourier series approach produces an accurate approximation at (-8, 8), and does not lead to gradient explosion as inputs go beyond the desired range. Although Fourier series is a well-known approach to approximate non-linear functions, it is not commonly used in the MPC literature as a function approximator mainly because of the difficulty to privately evaluate the $\mathsf{Sin}$ function. The $\mathsf{Sin}$ function itself is typically evaluated using some function approximators. Central to our approach is a novel protocol $\Pi_{\mathsf{Sin} _m}$  that can privately evaluate $\sin \frac{2 k \pi x }{2^m}$, $m\in \mathbb{Z}$ in just a single round. We rely on the following equation
\[
\sin \frac{2 k \pi x }{2^m} = \sin \frac{2 k \pi (x-\tilde{x}) }{2^m} \cos \frac{2 k \pi \tilde{x} }{2^m} + \cos \frac{2 k \pi (x-\tilde{x}) }{2^m} \sin \frac{2 k \pi \tilde{x} }{2^m} 
\]
to mask the input $x$ by $\tilde{x}$ and split $\sin \frac{2 k \pi x }{2^m}$ into shares.  The full protocol is described in Algorithm \ref{alg:Sin}.




\begin{algorithm}[htbp] \footnotesize
  \caption{Protocol $\Pi_{\mathsf{Sin} _m} $  }
  \label{alg:Sin}
  \begin{algorithmic}[1]
    \REQUIRE
      $P_0, P_1$ hold  shares of $x$ over $\frac{1}{2^f} \mathbb{Z}/2^n\mathbb{Z}$, and a public vector $k \in \mathbb{Z}$ 
    \ENSURE
      $P_0$, $P_1$ get the shares of $\left[ \sin \frac{2 k \pi x }{2^m} \right] _{\frac{1}{2^f}\mathbb{Z} }$ over $\frac{1}{2^f}\mathbb{Z}/2^{n-f}\mathbb{Z}$.
 \STATE $P_2$ generates random $\tilde{x}_L, \tilde{x_R} \in \frac{1}{2^f}\mathbb{Z}/2^m\mathbb{Z}$, $u_L, v_L \in \frac{1}{2^f}\mathbb{Z}/2^{n-f}\mathbb{Z}$,  and compute $\tilde{x}:=\tilde{x}_L + \tilde{x_R} \in \frac{1}{2^f}\mathbb{Z}/2^m\mathbb{Z}$, $u_R= \left[ \sin \frac{2 k \pi \tilde{x} }{2^m} \right] _{\frac{1}{2^f}\mathbb{Z} } -u_L \in  \frac{1}{2^f}\mathbb{Z}/2^{n-f}\mathbb{Z}, v_R = \left[ \cos \frac{2 k \pi \tilde{x} }{2^m} \right] _{\frac{1}{2^f}\mathbb{Z} } -v_L \in \frac{1}{2^f}\mathbb{Z}/2^{n-f}\mathbb{Z}$.
 \STATE $P_2$ sends $\tilde{x}_L, \tilde{u}_L, \tilde{v}_L$ to $P_0$ and sends $\tilde{x}_R, \tilde{u}_R, \tilde{v}_R$ to $P_1$; 
 \STATE $P_0$ computes $\delta x_L :=x _L -\tilde{x}_L \in \frac{1}{2^f}\mathbb{Z}/2^m\mathbb{Z}$ and send to $P_1$; 
 \STATE $P_1$ computes $\delta x_R:=x_R - \tilde{x}_R \in \frac{1}{2^f}\mathbb{Z}/2^m\mathbb{Z}$ and send to $P_0$;
 \STATE $P_0, P_1$ reconstruct $\delta x:=\delta x_L+\delta x_R \in \frac{1}{2^f}\mathbb{Z}/2^m \mathbb{Z}$, and compute $s:=\sin \frac{2 k \pi \delta x }{2^m} \in  \frac{1}{2^f}\mathbb{Z}/2^{n-f}\mathbb{Z}, c:= \cos \frac{2 k \pi \delta x }{2^m} \in \frac{1}{2^f}\mathbb{Z}/2^{n-f}\mathbb{Z}$ respectively
    \STATE $P_0$ compute $w_L:=sv_L+cu_L \in \frac{1}{2^{2f}}\mathbb{Z}/2^{n-2f}\mathbb{Z}$, $P_1$ compute $w_R:=sv_R+cu_R \in \frac{1}{2^{2f}}\mathbb{Z}/2^{n-2f}\mathbb{Z}$
    \STATE Return $(w_L, w_R)$
  \end{algorithmic}
\end{algorithm}

 \textbf{Analysis of Communication Complexity}. We denote $n$ as the bit length of the chosen ring, $f$ as the fixed-point precision, and $m$ the positive integer in $\sin \frac{2 k \pi x }{2^m}$. Step 1 and 2 of $\Pi_{\mathsf{Sin} _m}$ can be computed offline, and all the following steps need to be executed during the online phase. The offline communication of protocol $\Pi_{\mathsf{Sin}}$ requires sending $\tilde{x}_L, \tilde{u}_L, \tilde{v}_L$ to $P_0$ and $\tilde{x}_R, \tilde{u}_R, \tilde{v}_R$ to $P_1$, which is $4n+2(m+f)$ bits. We can further reduce this communication to $2n$ by using PRFs to generate $\tilde{x}_L, \tilde{x}_R, \tilde{u}_L, \tilde{v}_L$ locally, and further reduce the communication to $2n$ bits. We leave the detail of this optimization to Appendix~\ref{append:sin}. During the online phase, the only communication comes from reconstructing $\delta x = \delta x_L + \delta x_R$, which requires only one round and $2(m+f)$ bits of communication.
 
 Evaluating sigmoid is equivalent to evaluating 5 $\sin \frac{2 k \pi x }{2^m}$ functions in parallel. It is worth noting that since the input $x$ remains the same in all these $\mathsf{Sin}$ functions, we can reuse $\delta x$ during the online phase. Therefore, the online communication cost does not scale with the number of $\mathsf{Sin}$ functions we evaluate. The detailed comparison of communication complexity of the sigmoid function is given in Table \ref{table:sigmoid_comm}. Our protocol reduces online communication significantly. Given $n=64, f=14, m=5$, the online communication of our protocol $\Pi_{\mathsf{Sigmoid}} ^ \mathsf{fourier}$ is only $3\%$ that of $\Pi_{\mathsf{Sigmoid}} ^ \mathsf{Poly}$ (the Chebyshev approximation)


 
 \begin{table}[htbp]  \footnotesize
\begin{center}
\begin{threeparttable}
\begin{tabular}{ccccc}
\hline
    & offline  & online  & online  & total \\
  Protocol & comm. & comm. & round & comm. \\
  \hline
$\Pi_{\mathsf{Sigmoid}} ^ \mathsf{poly}$ &  5n  &  20n   & 5 &  25n \\
\hline
$\Pi_{\mathsf{Sin}}$ &  2n  & 2(m+f)    & 1 &  2(n+m+f)\\
$\Pi_{\mathsf{Sigmoid}} ^ \mathsf{fourier}$  &  10n  & 2(m+f)  & 1 &  2(5n+m+f)  \\
\hline
\label{T_sigmoid}
\end{tabular}
    \end{threeparttable}
\caption{Communication Complexity of $\Pi_{\mathsf{Sigmoid}}$ and $\Pi_{\mathsf{Sin}}$}
\label{table:sigmoid_comm}
\end{center}
\end{table}

\subsection{Softmax}
The function $softmax: \mathbb{R}^m \longrightarrow \mathbb{R}^m$ is defined as
\[
softmax(x)_i:= \frac{\exp (x_i) }{ \sum _i \exp (x_i) }
\]
It is an essential building block for multi-class classification tasks in machine learning. Given an input vector, the softmax function transforms it into a normalized probability score vector. We can then apply certain loss functions on the probability scores and train a machine learning model. Since the softmax function needs to compute the exponential function, which may potentially overflow when the input gets large, people in practice often first subtract the input vector by the maximum element in the vector before applying the softmax function. 
\[
\begin{array}{rl}
x_m := & \max _i \{x_i\} \\
softmax(x)  := & \frac{\exp(x_i-x_m)}{\sum _i \exp (x_i-x_m)}
\end{array}
\]

Most prior work on privacy-preserving machine learning focuses on private inference and does not add support for computing the softmax function. A few work~\cite{SecureNN, Falcon} that support private training replaces the exponential function in softmax by the ReLU function. This approach is both expensive in communication and lowers training accuracy. CrypTen uses the standard two step processes to compute softmax, which first evaluates maximum privately, and then privately evaluate exponentiation and division. Although this approach is able to produce accurate outputs, this two-step process is known to be expensive in communication complexity. Assuming the vector size is $m$, the private evaluation of maximum requires $\log m$ rounds of private comparison, which itself would take $O(\log n)$ rounds. (n is the bit length of the ring). Both exponentiation and private division requires around 10 rounds each to approximate. To evaluate a softmax function on 64 classes, the protocol would already take roughly 60 rounds of communication, and the cost further scales as we increase the class size. The cost of evaluating the softmax function accurately is huge!

We present our novel protocol to approximate the softmax function efficiently, requiring only 32 rounds of communication, and much less communication traffic. Moreover, the communication round does not scale with the class size. Our protocol produces accurate approximation when the class size scales to 10000, and the round complexity remains fixed. Our approach is based on the initial value problem (ivp) for ordinary differential equation~\cite{ode}. As far as we know, we are the first work to employ such technique in secure multi-party computation. 

\paragraph{Protocol Description}. Let $f: [0,1] \longrightarrow R^m$ defined by $f(t)=softmax(tx)$. Then we have
\[
\begin{array}{lcl}
f(0) & = & 1/m \\
f'(t) & =  & (x - \langle x, f(t) \rangle 1_m ) * f(t)
\end{array}
\]
where 
$1_m$ means the  vector $(1, \cdots 1)^T$ of dimension $m$, 
$\langle x, y \rangle$ means the inner value of vector $x$ and $y$,  $*$ means element-wise multiplication.

We wish get $f(1)=softmax(x)$. This is an Initial-Value Problems for  ODE (Ordinary Differential Equations).
We use the Euler formula to solve this ODE
\begin{equation}
\label{softmax_iter_formula}
\begin{array}{lcl}
y_0 & = & 1_m /m \\
y_{t+1} & = & y_t+ (x - \langle x, y_t \rangle 1_m ) * y_t/k
\end{array}
\end{equation}
for $t=0, \cdots, k-1$. Finally, we get $softmax(x)=y_k$. The formal description of protocol $\Pi_{\mathsf{Softmax}}$ is given below.

\begin{algorithm}[htbp] \footnotesize
  \caption{Protocol $\Pi_{\mathsf{Softmax}} $  }
  \label{alg:Softmax}
  \begin{algorithmic}[1]
    \REQUIRE
      $P_0, P_1$ hold  shares of a vector $x$ of dimension m
    \ENSURE
      $P_0$, $P_1$ get the shares of softmax$(x)$.
\STATE $P_0, P_1$ set shares of $y_0=1_m/m$
\FOR {$t = 0 \cdots k-1$}
 \STATE $P_0, P_1$ compute the shares of  $z=x * y_t$
 \STATE $P_0, P1$ compute the shares of $y_{t+1}:=y_t+(z-sum(z)*y_t)/k$ 
    \ENDFOR 
 \STATE $P_0, P_1$ Return the shares of $y_k$
  \end{algorithmic}
\end{algorithm}

\textbf{Analysis of Communication complexity}. Noted the formula  can be write as
\[
\begin{array}{lcl}
z_t & = & x * y_t \\
y_{t+1} & = & y_t + (z_t - sum(z_t) * y_t) /k
\end{array}
\]
We denote $m$ as the class size and $n$ as the bit length of the ring. Each iteration requires running beaver multiplication protocols twice and hence takes two rounds. For the first multiplication, the mask for $x$ can be reused to save communication. We only need to communicate the mask of $y_t$ in each iteration, and the mask of $x$ only once during the first iteration. The amortized communication cost is $2mn$ bits. For the second multiplication, $sum(z_t)$ is a scalar secret share and $y_t$ is a vector of shares. The mask of $y_t$ from the previous multiplication can be reused. We only need to communicate the mask of $sum(z_t)$, which takes $2n$ bits. The total per iteration online cost is $2(m+1)n$ bits in 2 rounds. With 16 iterations, the total communication cost is $32(m+1)n+2n$ bits.

%% file: experiment.tex
\section{Experiments}
All our experiments are conducted on three Linux CentOS servers, each equipped with an Intel(R) Xeon(R) Platinum 8163 CPU @ 2.50GHz CPU and 16GB memory, and in a WAN network setting. The network bandwidth is 30.1MB/s and the network delay is measured to be 50ms. To investigate the performance improvement of our protocol, we conduct experiments in the following two settings: logistic regression for binary classification, and convolutional neural network for multi-class image recognition problem. We also run microbenchmarks to further investigate the scalability and accuracy of our protocols.

\subsection{Logistic Regression for Binary Classification}
Logistic regression is a widely used machine learning algorithm in the industry and academia for the task of binary classification. We evaluate the performance of logistic regression on the credit10 dataset, which is a subset of the open dataset GiveMeSomeCredit \cite{credit}. The dataset contains 8429 examples in the training set and 3614 examples in the testing set, and each example has 10 features. We compare our system to other three PPML framework that also supports logistic regression training (TFE, Rosetta, CrypTen) and present our result in Table \ref{table:lr}. We set the batch size to 128 in all of these experiments. 

From Table \ref{table:lr}, our system is 1.83x faster than prior systems for logistic regression training. This speedup is obtained relative to TFE since Rosetta produce off-the-chart KS and AUC scores. Our system also achieves the highest KS (Kolmogorov–Smirnov) and AUC (Area Under Curve) scores that is comparable with plaintext training results. 

\begin{table}[htbp] 
\begin{center}
\begin{threeparttable}
\begin{tabular}{ccccccc}
\hline
\hline
System & KS & AUC & Time (s) & Speed up \\
\hline
Rosetta & 0.0 & 0.500 & 0.417 & - 
\\
TFE & 0.475 & 0.801 & 0.565 & - 
\\
CrypTen & 0.496 & 0.826 & 35.1 & -
\\
STF (ours) & \textbf{0.496} & \textbf{0.826} & \textbf{0.309} & {\color{red}1.83x} \tnote{*} \\
\hline
Plaintext & 0.496 & 0.825 & - & - \\
\hline
\end{tabular}
\vskip15pt
\begin{tablenotes}
        \footnotesize
        \item[*] Since Rosetta's KS and AUC scores indicate that it does not train correctly, this speedup is relative to TFE. 
        \end{tablenotes}
    \end{threeparttable}
\caption{Runtime (in seconds), KS, AUC of for a single iteration of privacy-preserving logistic regression training on the Credit10 Dataset, with a batch size of 128.} \label{table:lr}. 
\end{center}
\end{table}

\subsection{Convolution Neural Network for Multi-class Classification}
For the multi-class classification problem, we evaluate several convolutional neural networks (CNN) on the MNIST ~\cite{MNIST} dataset. We consider Network B to Network D described in SecureNN~\cite{SecureNN} for performance comparison. All these three networks are convolutional neural network. We use the cross-entropy loss as the loss function, which is the standard option for multi-class classification problem. To compute the cross entropy loss, a softmax function added on top of the network to transform the network output into a probability distribution.


We compare our system with SecureNN~\cite{TFE} and CrypTen \cite{crypten2020} that supports training these networks and present the result in the following Table \ref{table:cnn}. We do not compare against TFE \cite{TFE}, Rosetta \cite{Rosetta} since they do not support private training of convolutional neural network. Each entry in the table represents the runtime (in seconds) to train a single batch of 128 images. From Table~\ref{table:cnn}, our system achieves a 3.9-4.9x speedup on these networks over prior systems.

\begin{table}[htbp]
\begin{center}
\begin{threeparttable}
\begin{tabular}{cccc}
\hline
\hline
Network & System & Time & Speed up \\
\hline
Network B & CrypTen & 106.69 & 
\\
Network B & SecureNN  &  48.02  & 
- \\
Network B & STF (ours) &  \textbf{12.24}  &  {\color{red} 3.92x}  \\
\hline
Network C & CrypTen & 131.86 &  
- \\
Network C & SecureNN & 76.51 &  
- \\
Network C & STF (ours) &  \textbf{16.95} & {\color{red}4.51x}  \\
\hline
Network D & CrypTen & 75.72 &  
- \\
Network D & SecureNN & 16.15  &  
- \\
Network D & STF (ours) & \textbf{3.32} & {\color{red}4.86x}  \\
\hline
\end{tabular}
    \end{threeparttable}
\caption{Runtime (in seconds) of a single iteration of privacy-preserving neural network training on the MNIST dataset, with a batch size of 128.}
\label{table:cnn}
\end{center}
\end{table}

\paragraph{Accuracy of private CNN training.} 

To demonstrate the robustness of our system, we compare the training accuracy of Network B to Network D on STF with Keras~\cite{keras}, a widely-used plaintext machine learning framework. (See Table \ref{table:acc}). We initialized each network with the Glorot uniform distribution \cite{glorot}, and train each network using Stochastic Gradient Descent (SGD)~\cite{sgd,sgd2} with a learning rate of 0.01. The batch size is set to 128. We record the training accuracy at the end of epoch 1 and epoch 5. From Table \ref{table:acc}, the training accuracy of STF is comparable to that of Keras. The results indicate that our system is able to train neural networks both accurately and privately.

Robustness of privacy-preserving systems eventually boils down to the robustness each individual protocols. The major source of error for privacy-preserving protocols comes from evaluating functions that require approximation (e.g. reciprocal, log, exp, softmax). For training neural networks, softmax is the only function that we approximate. In Section \ref{sec:microbenchmarks}, we further explore the scalability and robustness of our softmax approximator. We show that our softmax protocols can produce accurate approximation even when we scale the class size to 10000.

\begin{table}[htbp]
\begin{center}
\begin{threeparttable}
\begin{tabular}{cccc}
\hline
\hline
Network & System  & Acc. at epoch 1 & Acc. at epoch 5\\
\hline
Network B & Keras  &  96.7\% & 98.7\% \\
Network B & STF &   96.4\% & 98.1\% \\
\hline
Network C & Keras &  97.3\% & 98.7\% \\
Network C & STF &   97.8\% & 98.5\% \\
\hline
Network D& Keras &96.5\%& 97.9\% \\
Network D& STF &  95.5\% & 98.2\% \\
\hline
\end{tabular}
\caption{Training accuracy (\%) comparisons of STF and Keras (plaintext) on the MNIST dataset, for Network B, C, D.}
\label{table:acc}
    \end{threeparttable}
\end{center}
\end{table}

\subsection{Microbenchmarks}
\label{sec:microbenchmarks}
\paragraph{Performance of Backward Convolution.} In Section \ref{sec:bilinear}, we extend the beaver triple protocol to enable efficient triple distribution for bilinear map. We further show that both the forward and backward propagation of convolution satisfy the properties of bilinear maps, and we can therefore apply the extended protocols on them. To demonstrate the performance improvement of the extended protocol, we evaluate the backward propagation on convolution filters with varying number of input channels (from 32 to 512). We fix the convolution kernel size to be $3\times3$, the input dimension to be $16\times 16$, and set the batch size to 128. We compare the runtime of STF, which leverages the extended protocol, with SecureNN and CrypTen which does not. 

\begin{figure}
\begin{tikzpicture}
\begin{axis}[
    xmode=log,
    log basis x={2},
    title={},
    xlabel={Number of input channels},
    ylabel={Time (s/batch)},
    xmin=0, xmax=512,
    ymin=0, ymax=120,
    xtick=data,
    ytick={0, 30, 60, 90, 120},
    legend pos=north west,
    xmajorgrids=true,
    ymajorgrids=true,
    grid style=dashed,
]
\addplot[
    color=blue,
    mark=square,
    ]
    coordinates {
    (32,7.226)(64,14.022)(128,27.589)(256,55.758)(512,111.023)
    };
\addplot[
    color=purple,
    mark=triangle,
    ]
    coordinates {
(32,2.72568)(64,5.69614)(128,11.4382)(256,23.2656)(512,46.5116)
    };
\addplot[
    color=red,
    mark=halfcircle,
    ]
    coordinates {
    (32,0.502)(64,0.792)(128,1.404)(256, 2.648)(512, 5.095)
    };

\legend{CrypTen, SecureNN, STF (ours)}
\end{axis}
\end{tikzpicture}
\caption{Runtime comparisons (in seconds) of backward propagation on convolution filters (kernels) with varying number of input channels. The convolution layer has 16 output channels, the kernel size is $3\times3$, the input dimension is $16\times 16$, and the batch size is 128.}
\label{fig:conv_back}
\end{figure}
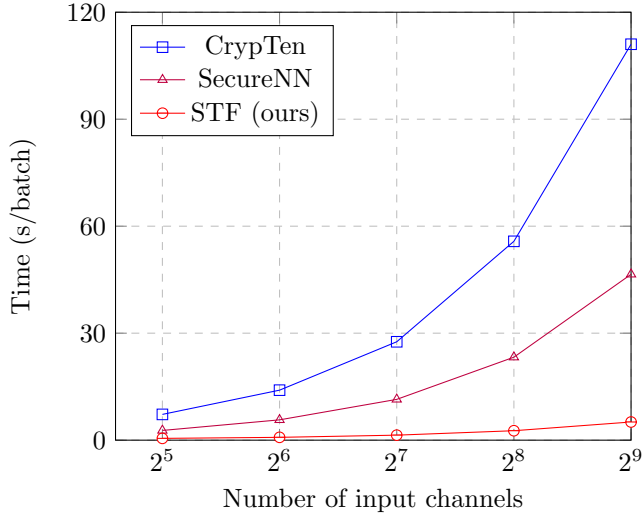

Results in Figure \ref{fig:conv_back} shows that the extended beaver triple protocols for bilinear maps translates to a significant performance gain. At 32 input channels, it takes STF 0.5 second and SecureNN 2.7 seconds to execute its protocol. The gap gets bigger as we increase the number of channels. At 512 input channels, STF takes 5.1 seconds to evaluate the protocol whereas SecureNN takes 46.5 seconds. Overall, STF achieves a 5-9x speedup over SecureNN and 11-21x over CrypTen on convolution's backward pass on filters.      

\paragraph{Robustness and Scalability of Softmax.} As mentioned in Section \ref{sec:non-linear}, the standard approach to evaluate softmax requires evaluating the combination of maximum, exponentiation, summation, and division operators. This is the approach used by CrypTen to evaluate softmax. SecureNN does not evaluate softmax directly, but uses the following function $ASM$ to replace softmax:
\[
ASM(x)_i = \frac{ReLU(x_i)}{\sum_i {ReLU(x_i)}}
\]
The above two approaches are both expensive to evaluate, since they require computing private comparison and division. In contrast, our method approximate softmax as a whole, and removes the need to run expensive private comparison and division protocols at all. Moreover, our protocol requires a constant 32 rounds to evaluate, which is independent of the class size. 

\begin{figure}
\begin{tikzpicture}
\begin{axis}[
    xmode=log,
    title={},
    xlabel={Number of classes},
    ylabel={Time (s)},
    xmin=0, xmax=20000,
    ymin=0, ymax=550,
    xtick={10, 100, 1000, 10000},
    ytick={0, 100, 200, 300, 400, 500},
    legend pos=north west,
    xmajorgrids=true,
    ymajorgrids=true,
    grid style=dashed,
]

\addplot[
    color=blue,
    mark=square,
    ]
    coordinates {
    (10,29.6)(100,39.6)(1000,51.5)(10000, 112.1)
    };

\addplot[
    color=purple,
    mark=triangle,
    ]
    coordinates {
(10,3.25565)(100, 3.89969)(1000, 49.4129)(10000,533.787)
    };

\addplot[
    color=red,
    mark=halfcircle,
    ]
    coordinates {
    (10,1.0)(100,1.14)(1000,2.93)(10000, 26.30)
    };

\legend{CrypTen, SecureNN, STF(ours)}
\end{axis}
\end{tikzpicture}
\caption{Runtime comparisons (in seconds) of the softmax function with varying number of classes. The batch size is set to 128.}
\label{fig:softmax_runti}
\end{figure}
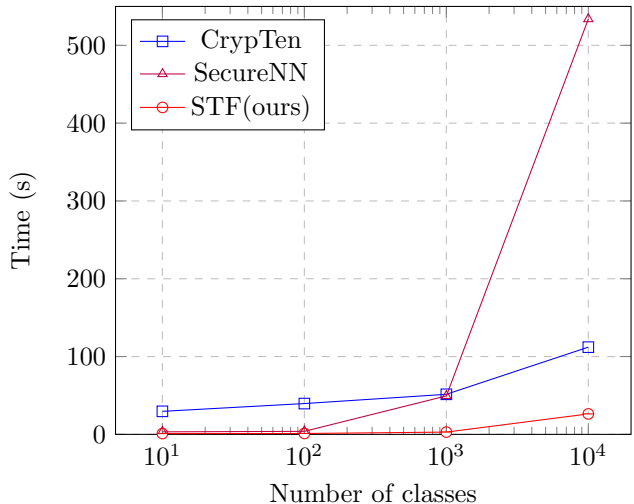

At Figure \ref{fig:softmax_runti}, we plot STF, SecureNN, and CrypTen's runtime of evaluating softmax under different class sizes. At 10 classes, which is the class size of MNIST, it takes STF only 1.0 second to evaluate softmax, whereas SecureNN and CrypTen requires 3.2 and 29.6 seconds respectively. At 1000 classes, the class size of ImageNet, it takes STF 2.93 seconds to evaluate softmax, a 17x speedup over SecureNN and CrypTen.

Our softmax protocol is also much more robust than protocols in CrypTen and SecureNN. At Table \ref{table:softmax-kl}, we record the KL divergence between the plaintext softmax function and the softmax approximators. Although CrypTen's softmax protocol has lower KL divergence scores at 10 and 100 classes, it produces off-the-chart results when the class size scales to 1000 and 10000. SecureNN's softmax protocol (ASM) does not produce an accurate approximation for Softmax, and therefore has a much higher KL divergence score. In comparison, our protocol provides a consistently accurate approximation of softmax even when the class size scales to 10000.  

\begin{table}[htbp] 
\begin{center}
\begin{tabular}{ccccccc}
\hline
System/Class size & 10 & 100 & 1000 & 10000 \\
\hline
Uniform & 0.3731 & 0.4842 & 0.5026 & 0.4992 \\
ASM &  0.063 & 0.068 & 0.07 & 0.07 \\
CrypTen & 8.6e-6 & 2.3e-5 & Nah & Nah \\
STF (ours) & 0.0003 & 0.0010 & 0.0015 & 0.0065  \\
\hline
\end{tabular}
\end{center}
\caption{KL Divergence between the plaintext softmax function and the softmax approximation function of STF and CrypTen. Inputs are sampled from a Gaussian distribution $\mathcal{N}(0,1)$. Uniform serves as a baseline algorithm that assumes the softmax output follows a uniform distribution. Each output element in Uniform is set to $1/n$ where n is the class size.}
\label{table:softmax-kl}
\end{table}


%% file: conclusion.tex
\section{Related Work}
\paragraph{Privacy-preserving machine learning.} There are a number of recent work that leverages MPC to build privacy-preserving ML systems. These works operate under a variety of security models. Some of these works consider a 2-party setting~\cite{SecureML, DeepSecure, EzPC, Gazelle,MiniONN, XONN, Delphi, ABY2, CrypTFlow2}, some works operate in and in a 3-party setting~\cite{ABY3, crypten2020,cryptflow, Falcon}, and others in the 4-party setting~\cite{Blaze,Flash}. Only a few of these works support private training. Although we consider the 3-party setting with semi-honest security, our protocols for sigmoid and softmax can nevertheless be easily transferred to other settings as well. 

\paragraph{Other related work.} Besides secure MPC, there are other techniques that can provide privacy guarantees for collaborative machine learning. Differential privacy~\cite{dp1,dp2} provides privacy by adding some noise to the model, and has been applied to train deep neural networks~\cite{dp_deep}. Another approach relies on hardware enclave such as Intel SGX~\cite{sgx} to provide a Trusted Execution Environment (TEE). Data and codes in TEE are isolated from the operating system, and no one without integrity can access the data or alter the code. Systems for privacy-preserving ML on TEE~\cite{tee1, tee2} has been researched and developed.

\section{Conclusion}
In this work, we present Morse-STF, a privacy-preserving machine learning system that achieves state-of-the-art performance in LR and CNN training. Our system leverages novel protocols for evaluating convolution, sigmoid, and softmax functions that significantly reduce communication of each these components. Our protocols for sigmoid and softmax are based on new approximation methods. These methods provide new tools to develop other non-linear protocols more efficiently. 


%% file: appendix.tex
\section{Appendix}

\subsection{Protocol for Convolution}
\label{append:conv}
We give our protocols for convolution and its backward propagation below. 

\begin{algorithm}[htbp] \footnotesize
  \caption{Protocol $\Pi_\mathsf{Conv2D}$  }
  \label{conv2d_ss}
  \begin{algorithmic}[1]
    \REQUIRE
      $P_0$, $P_1$ hold the shares of  input over $\bigotimes _{B, m, n, C} A$, and the shares of filter over $\bigotimes _{C, r, s, D} A$.
      
       $\frac{\partial loss}{\partial output} =\frac{\partial loss}{\partial output} _0+\frac{\partial loss}{\partial output}  _1 \in \bigotimes _{B, m', n', D} A$,  $filter=filter_0+filter_1 \in  \bigotimes _{C, r, s, D} A $.
    \ENSURE
      $P_0$, $P_1$ get the shares of output over $\bigotimes _{B, m', n', D} A$.
 \STATE $P_0$, $P_1$ call the protocol $\Pi_{\mathsf{BM}}$ to compute output.
  \end{algorithmic}
\end{algorithm}

\begin{algorithm}[htbp] \footnotesize
  \caption{Protocol $\Pi_{\mathsf{Con2D_{bi}}}$  }
  \label{protocol_conv2d_bi}
  \begin{algorithmic}[1]
    \REQUIRE
      $P_0$, $P_1$ hold the shares of  $\frac{\partial loss}{\partial output} =\frac{\partial loss}{\partial output} _0+\frac{\partial loss}{\partial output}  _1 \in \bigotimes _{B, m', n', D} A$,  $filter=filter_0+filter_1 \in  \bigotimes _{C, r, s, D} A $.
    \ENSURE
      $P_0$, $P_1$ get the shares of $\frac{\partial loss}{\partial input} $ over $\bigotimes _{B, m, n, C} A$.
 \STATE $P_0$, $P_1$ call the protocol $\Pi_{\mathsf{BM}}$ to compute $\frac{\partial loss}{\partial input}$.
  \end{algorithmic}
\end{algorithm}

\begin{algorithm}[htbp]  \footnotesize
  \caption{Protocol $\Pi_{\mathsf{Con2D_{bf}}}$  }
 \label{protocol_conv2d_bf}
  \begin{algorithmic}[1]
    \REQUIRE
      $P_0$, $P_1$ hold the shares of  $\frac{\partial loss}{\partial output} =\frac{\partial loss}{\partial output} _0+\frac{\partial loss}{\partial output}  _1 \in \bigotimes _{B, m', n', D} A$,  $input= input_0 + input _1 \in \bigotimes _{B, m, n, C} A$.
    \ENSURE
      $P_0$, $P_1$ get the shares of $\frac{\partial loss}{\partial filter} $ over $ \bigotimes _{C, m, n, D} A$.
 \STATE $P_0$, $P_1$ call the protocol $\Pi_{\mathsf{BM}}$ to compute $\frac{\partial loss}{\partial filter}$.
  \end{algorithmic}
\end{algorithm}

\subsection{Proof of Theorem \ref{theorem:back_prop}}
\textbf{Proof of Theroem \ref{theorem:back_prop}:}\\
a. 
We can write the map $f$ as 
\[
z_k=\sum _{i, j} a_{i,j}^k x_i y_j
\]
For any fixed $i$, we have
\[
\frac{\partial loss}{\partial x_i} =  \sum _k \frac{\partial loss}{\partial z_k} \frac{\partial z_k}{\partial x_i} =\sum _k \sum _j p_{k,j} \frac{\partial loss}{\partial z_k} y_j 
\]
where
\[
p_{k,j} = a_{i,j}^k 
\]
Hence $\frac{\partial loss}{\partial x_i}$ is bilinear of $(\frac{\partial loss}{\partial z_k})_k$ and $y$. 

b. The proof is similar to the proof of a. \qed

\subsection{Communication complexity of convolution and its back propagation}
We describe the communication complexity of convolution and its back propagation in Table  \ref{T_conv2D}.
We compare the  communication complexity of convolution and its back propagation to other works in Table  \ref{comm_conv2D_example}.
\begin{table*}[htbp] 
  \begin{center}
  \begin{threeparttable}
  \begin{tabular}{ccccc}
  \hline
    Protocol  & Offline Comm.  & Online Comm.  & Online Round & Total Comm.\\ \hline
  $\Pi_{\mathsf{Conv2D}} ^{}$ & $(m'n'BD)\log|A|$ & $2(mnBC+rsCD)\log |A|$ & 1 & $(2mnBC+2rsCD+m'n'BD)\log |A|$ \\
  $\Pi_{\mathsf{Conv2D_{bi} } } ^{}$ & $(mnBC)\log |A|$ & $2(m'n'BD+rsCD)\log |A|$ & 1 & $(mnBC+2rsCD+2m'n'BD)\log |A|$ \\
  $\Pi_{\mathsf{Conv2D_{bf}}} ^{}$ & $(rsCD)\log |A|$ & $2(mnBC+m'n'BD)\log |A|$ & 1 & $(2mnBC+rsCD+2m'n'BD)\log |A|$ \\
  \hline
  \end{tabular}
  \end{threeparttable}
  \end{center}
  \caption{Communication complexity of $\Pi_{\mathsf{Conv2D}}, \Pi_{\mathsf{Conv2D_{bi}}}, \Pi_{\mathsf{Conv2D_{bf}}}$ for Morse-STF. $m, n$ is the height and width of the input, B is the batch size, C is the number of input channel, and D is the number of output channel. $r$, $s$ is the height and width of the filter (kernel), and $m'$, $n'$ is the height and width of the output. $|A|$ represents the size of the ring.}
  \label{T_conv2D}
  \end{table*}

  \begin{table*}[htbp] 
    \begin{center}
    \begin{threeparttable}
    \begin{tabular}{ccccc}
    \hline
      Protocol  & Offline Comm. (MB)  & Online Comm. (MB)  &  Total Comm. (MB)  & Savings \\ \hline

    $\Pi_{\mathsf{Conv2D}}$ in SecureNN & 3.13 &  62.88 &   66.01 & - \\
    $\Pi_{\mathsf{Conv2D}}$ in CrypTen, STF (ours) & 3.13 & 6.01 &  9.13  & 86.17\%\\
    \hline 
    $\Pi_{\mathsf{Conv2D_{bi} } }$ in SeureNN & 2.81 & 351.94  & 354.76 & - \\
    $\Pi_{\mathsf{Conv2D_{bi} } }$ in CrypTen, STF (ours) & 2.81 & 6.63 & 9.44 & 97.34\% \\
    \hline 
    $\Pi_{\mathsf{Conv2D_{bf}}} $ in CrypTen & 24.41 & 130.63 &  155.04 & -\\
    $\Pi_{\mathsf{Conv2D_{bf}}} $ in SecureNN & 0.19 & 68.75 &  68.94 & -\\
    $\Pi_{\mathsf{Conv2D_{bf}}} $ in STF (ours) & 0.19 &  11.88 &  12.07 & 82.49\% \\
    \hline
    \end{tabular}
    \end{threeparttable}
    \end{center}
    \caption{Communication complexity comparison of $\Pi_{\mathsf{Conv2D}}, \Pi_{\mathsf{Conv2D_{bi}}}, \Pi_{\mathsf{Conv2D_{bf}}}$ between CrypTen, SecureNN and STF (ours). We use the second convolution layer of Network C as an example. The input shape is (B, m, n, C) = (128, 12, 12, 20), filter shape is (C, r, s, D) = (20, 5, 5, 50), and the corresponding output shape is (B, m', n', D) = (128, 8, 8, 50).}
    \label{comm_conv2D_example}
  \end{table*}

\subsection{Protocol for $\mathsf{Sin}$ with PRF}
We describe the Protocol for $\mathsf{Sin}$ with PRF in Algorithm \ref{alg:Sin_PRF}.
\label{append:sin}
\begin{algorithm}[htbp] \footnotesize
  \caption{Protocol $\Pi_{\mathsf{Sin} _m} $  }
  \label{alg:Sin_PRF}
  \begin{algorithmic}[1]
    \REQUIRE
      $P_0, P_1$ hold shares of $x$ over $\frac{1}{2^f} \mathbb{Z}/2^n\mathbb{Z}$, and a public vector $k \in \mathbb{Z}$; $P_0, P_2$ hold PRF$_0$ with key $k_0$, $P_1, P_2$ hold PRF$_1$ with key $k_1$.
    \ENSURE
      $P_0$, $P_1$ get the shares of $\left[ \sin \frac{2 k \pi x }{2^m} \right] _{\frac{1}{2^f}\mathbb{Z} }$ over $\frac{1}{2^f}\mathbb{Z}/2^{n-f}\mathbb{Z}$.
 \STATE $P_2$ generates $\tilde{x}_L \in \frac{1}{2^f}\mathbb{Z}/2^m\mathbb{Z}$, $u_L, v_L \in \frac{1}{2^f}\mathbb{Z}/2^{n-f}\mathbb{Z}$ using PRF$_0$; generate
 $\tilde{x}_R \in \frac{1}{2^f}\mathbb{Z}/2^m\mathbb{Z}$
 using PRF$_1$;
 and compute $\tilde{x}:=\tilde{x}_L + \tilde{x_R} \in \frac{1}{2^f}\mathbb{Z}/2^m\mathbb{Z}$, $u_R= \left[ \sin \frac{2 k \pi \tilde{x} }{2^m} \right] _{\frac{1}{2^f}\mathbb{Z} } -u_L \in  \frac{1}{2^f}\mathbb{Z}/2^{n-f}\mathbb{Z}, v_R = \left[ \cos \frac{2 k \pi \tilde{x} }{2^m} \right] _{\frac{1}{2^f}\mathbb{Z} } -v_L \in \frac{1}{2^f}\mathbb{Z}/2^{n-f}\mathbb{Z}$.
 \STATE $P_0$ generates $\tilde{x}_L \in \frac{1}{2^f}\mathbb{Z}/2^m\mathbb{Z}$, $u_L, v_L \in \frac{1}{2^f}\mathbb{Z}/2^{n-f}\mathbb{Z}$ using PRF$_0$; $P_1$ generate
 $\tilde{x}_R \in \frac{1}{2^f}\mathbb{Z}/2^m\mathbb{Z}$
 using PRF$_1$; $P_2$ sends $\tilde{u}_R, \tilde{v}_R$ to $P_1$; 
 \STATE $P_0$ computes $\delta x_L :=x _L -\tilde{x}_L \in \frac{1}{2^f}\mathbb{Z}/2^m\mathbb{Z}$ and send to $P_1$; 
 \STATE $P_1$ computes $\delta x_R:=x_R - \tilde{x}_R \in \frac{1}{2^f}\mathbb{Z}/2^m\mathbb{Z}$ and send to $P_0$;
 \STATE $P_0, P_1$ reconstruct $\delta x:=\delta x_L+\delta x_R \in \frac{1}{2^f}\mathbb{Z}/2^m \mathbb{Z}$, and compute $s:=\sin \frac{2 k \pi \delta x }{2^m} \in  \frac{1}{2^f}\mathbb{Z}/2^{n-f}\mathbb{Z}, c:= \cos \frac{2 k \pi \delta x }{2^m} \in \frac{1}{2^f}\mathbb{Z}/2^{n-f}\mathbb{Z}$ respectively
    \STATE $P_0$ compute $w_L:=sv_L+cu_L \in \frac{1}{2^{2f}}\mathbb{Z}/2^{n-2f}\mathbb{Z}$, $P_1$ compute $w_R:=sv_R+cu_R \in \frac{1}{2^{2f}}\mathbb{Z}/2^{n-2f}\mathbb{Z}$
    \STATE Return $(w_L, w_R)$
  \end{algorithmic}
\end{algorithm}